\documentclass[11pt]{article}
\usepackage{epsfig} 
\newcommand{\sect}[1]{ \section{#1} \setcounter{equation}{0} }

\newcommand{\half}{\mbox{\small{$\frac{1}{2}$}}}

\newcommand{\MSbar}{\overline{\mbox{MS}}} 

\begin{document}
\title{Critical exponent $\omega$ at $O(1/N)$ in $O(N) \times O(m)$ spin 
models} 
\author{J.A. Gracey, \\ Theoretical Physics Division, \\ Department 
of Mathematical Sciences, \\ University of Liverpool, \\ P.O. Box 147, \\ 
Liverpool, \\ L69 3BX, \\ United Kingdom.} 
\date{} 
\maketitle 
\vspace{5cm} 
\noindent 
{\bf Abstract.} We compute the $O(1/N)$ correction to the stability critical
exponent, $\omega$, in the Landau-Ginzburg-Wilson model with $O(N)$ $\times$ 
$O(m)$ symmetry at the stable chiral fixed point and the stable direction at 
the unstable antichiral fixed point. Several constraints on the $O(1/N)$ 
coefficients of the four loop perturbative $\beta$-functions are computed.  

\vspace{-15.5cm} 
\hspace{13.5cm} 
{\bf LTH 560} 

\newpage 

\sect{Introduction.}
The renormalization group analysis of spin models has proven to be an 
important tool in understanding and predicting critical properties of phase
transitions in a variety of phenomena. For a recent review see, for example,
\cite{1}. For instance, the Heisenberg model based on an $O(3)$ nonlinear sigma
model or $O(3)$ $\phi^4$ theory has been widely used to understand 
ferromagnetic phase transitions in nature. Indeed the perturbative field theory
techniques in this instance have been developed to five loops in $\MSbar$ in 
standard $d$-dimensional regularization in $\phi^4$ theory, \cite{2}, and 
equally impressively to six loops in the three dimensional fixed dimension 
renormalization, \cite{3,4,5}. Given such success these methods have been 
applied to similar models of other critical phenomena. Over a period of years 
various calculations have been carried out in the Landau-Ginzburg-Wilson model 
which essentially is an extension of $\phi^4$ theory where the $O(N)$ symmetry
is replaced by an $O(N)$ $\times$ $O(2)$ symmetry, \cite{6,7,8,9,10,11}. The 
critical structure in this generalised model is richer than the usual $\phi^4$ 
theory in that theoretically there are several fixed points over and above the 
Heisenberg one, depending on the value of $N$, which are either stable or 
unstable, \cite{10,11}. However, the properties of the phase transitions in  
this Landau-Ginzburg-Wilson class of models is controversial, 
\cite{9,11,12,13,14,15,16,17,18,19,20,21}. First, the two dimensional nonlinear
sigma model with the same symmetry is believed to reside in the same 
universality class, \cite{10,11}. Therefore, it ought to be possible to use 
either model to compute useful information on the critical properties of the 
physically interesting stable chiral fixed point. However, it has been pointed 
out in \cite{12} that the results from a $d$ $=$ $2$ $+$ $\varepsilon$ 
dimensional study do not match those deduced from the higher dimensional 
theory. Second, in field theoretical calculations such as perturbation theory 
the phase transitions are regarded as second order whilst numerical or Monte 
Carlo simulations would appear to indicate transitions are first order in 
nature, \cite{9,13,14,15,16,17,18}. Moreover, the particular behaviour depends 
on the value of $N$ though the precise range where this occurs is still 
undetermined. Further, one recent work has suggested an interesting point of 
view for the origin of the disagreements for $N$ $=$ $2$ and $3$ in both two 
and three dimensions. In \cite{22} it is argued that it is due to the fact that
one flows to the stable chiral fixed point along a spiral-like trajectory in 
contrast to the usual descent. To endeavour to clarify the issue the 
perturbative analysis of the Landau-Ginzburg-Wilson critical behaviour has 
recently been extended to a higher loop order in \cite{11}. Previous one and 
two loop computations were carried out in \cite{6,7,8}. The new three loop 
$\MSbar$ calculations of the renormalization group functions such as the 
$\beta$-functions of both coupling constants and anomalous dimensions, 
\cite{11}, have provided more accurate information on the fixed point locations
and the range of parameters for which they exist and are stable or not. Indeed 
in this respect the models with the more general symmetry group of $O(N)$ 
$\times$ $O(m)$ were studied with $m$ only set to $m$~$=$~$2$ at the end, 
\cite{11}. Such three loop calculations represent the current perturbative 
status of the model. 

However, it is in principle possible to extend this three loop $\MSbar$ 
dimensionally regularized calculation to the next order, though it will involve
a huge number of Feynman diagrams. In previous work in the simpler $O(N)$ 
models the perturbative computations were complemented with large $N$ 
calculations of the same renormalization group functions to several orders in 
powers of $1/N$. For instance, various critical exponents are available at both 
$O(1/N^2)$ and $O(1/N^3)$, \cite{23,24,25,26}, as functions of $d$, with $2$
$<$ $d$ $<$ $4$, which correspond through the critical renormalization group 
equation with the renormalization group functions. More correctly these
critical exponents were computed in $d$-dimensions at the non-trivial 
Wilson-Fisher fixed point of the $d$-dimensional $\beta$-function which
corresponds to the Heisenberg fixed point in three dimensions in $O(N)$ 
$\phi^4$ theory or the $O(N)$ nonlinear sigma model which does lie in the same 
universality class. The coefficients of the powers of $\epsilon$ in the 
$\epsilon$-expansion of such exponents in $d$ $=$ $4$ $-$ $2\epsilon$ 
dimensions are in exact agreement with the perturbative coefficients in the 
renormalization group function to the perturbative order they are known, at a 
particular order in $1/N$. More significantly the large $N$ critical exponents 
contain {\em new} higher order information in the uncomputed coefficients which
would therefore assist future perturbative calculations. Indeed in the 
Landau-Ginzburg-Wilson context various critical exponents have already been 
computed in the model with $O(N)$ $\times$ $O(m)$ symmetry at the two 
non-trivial fixed points which exist in addition to the Heisenberg fixed point,
\cite{6,11}. These are known as the chiral stable, (CS), and antichiral 
unstable, (AU), fixed points. Moreover, the results for the critical exponents 
$\eta$ and $\nu$ at $O(1/N^2)$ at both CS and AU are in agreement with the new 
perturbative results of \cite{11}. However, to extract the {\em new} 
information encoded in the exponents at four and higher loops in these 
$d$-dimensional functions in relation to the {\em four} dimensional theory one 
requires knowledge of the location of the fixed points to the same loop and 
large $N$ order. From the critical renormalization group equation such 
information is encoded in the critical exponent $\omega$ which relates to the 
critical $\beta$-function slope of the model in the universality class which is
renormalizable in four dimensions. In the Landau-Ginzburg-Wilson model this has
not yet been computed at $O(1/N)$ at either CS or AU. Therefore, it is the 
purpose of this article to rectify this gap and thereby unlock the door to 
higher order information on the structure of the perturbative renormalization 
group equations such as the $\beta$-function. In $O(N)$ $\phi^4$ theory this 
problem has already been resolved at $O(1/N^2)$, \cite{26}, where the $O(1/N)$ 
value for $\omega$ is relatively trivial to establish, \cite{27}, with the 
elegant machinery of the large $N$ critical point method of \cite{23,24}. 
However, for the CS and AU Landau-Ginzburg-Wilson fixed points the leading 
order, $O(1/N)$, analysis is much more involved since one is studying a model 
with two independent coupling constants. Therefore, whilst our calculation also
opens the road to an $O(1/N^2)$ computation, it represents a non-trivial 
example of how one treats the large $N$ formalism for $\omega$ exponents 
explicitly in a quantum field theory with more than one coupling constant which
deserves detailed treatment. 

The paper is organised as follows. In section two we recall the background
details of the model we are interested in and derive explicit expressions for
the location of the various fixed points from the explicit three loop 
perturbative results at $O(1/N)$ as well as the perturbative values of the 
eigenexponents of the stability matrix at criticality. These are related to the
exponents $\omega$ which we are interested in. Section three is devoted to the 
development of the large $N$ formalism for computing these various $\omega$ and
the explicit $d$-dimensional expressions are given at $O(1/N)$. Various 
concluding remarks are given in section four. 

\sect{Background.} 
The lagrangian of the massless Landau-Ginzburg-Wilson model involves a scalar
field with two quartic self interaction terms with an $O(N)$ $\times$ 
$O(m)$ symmetry and is given by  
\begin{equation} 
L ~=~ \frac{1}{2} (\partial_\mu \phi^{ai})^2 ~+~ \frac{u}{4!} \left( \phi^{ai} 
\phi^{ai} \right)^2 ~+~ \frac{v}{4!} \left[ \left( \phi^{ai} \phi^{bi} 
\right)^2 ~-~ \phi^{ai} \phi^{ai} \phi^{bj} \phi^{bj} \right] 
\label{lag} 
\end{equation}  
where $1$ $\leq$ $i$ $\leq$ $N$, $1$ $\leq$ $a$ $\leq$ $m$, and $u$ and $v$ are
the bare coupling constants. As in \cite{11} we rewrite (\ref{lag}) in order to 
perform the large $N$ expansion. This involves introducing two auxiliary scalar
fields one of which, $T^{ab}$, is a symmetric traceless tensor under $O(m)$. 
Thus (\ref{lag}) is equivalent to  
\begin{equation} 
L ~=~ \frac{1}{2} (\partial_\mu \phi^{ai})^2 ~+~ \frac{1}{2} \sigma \phi^{ai}
\phi^{ai} ~+~ \frac{1}{2} T^{ab} \phi^{ai} \phi^{bi} ~-~ 
\frac{3\sigma^2}{2w} ~-~ \frac{3}{2v} T^{ab} T^{ab}  
\label{lagcon} 
\end{equation} 
where $w$ $=$ $u$ $+$ $(m-1)v/m$ in our notation. If one uses the equations of
motion for $\sigma$ and $T^{ab}$ then (\ref{lag}) is recovered. The coupling
constants are defined in the kinetic terms to ensure that the vertices are in
the right form for applying the uniqueness technique to compute the large $N$
Feynman diagrams \cite{23,28}. One can understand the fixed point structure of 
(\ref{lag}) by considering the $\beta$-functions for each coupling which have 
been computed to several orders, \cite{8,11}. At three loops these are  
\begin{eqnarray} 
\beta_u(u,v) &=& \frac{1}{2}(d-4)u ~+~ \frac{(mn+8)}{6}u^2 ~-~ 
\frac{1}{3}(m-1)(n-1) v \left(u-\frac{v}{2} \right) ~-~ \frac{1}{6} (3mn+14) 
u^3 \nonumber \\ 
&& +~ (m-1)(n-1) \left( \frac{11}{9}u^2 - \frac{13}{12} uv + \frac{5}{18}v^2
\right) v \nonumber \\  
&& +~ \left[ [ 33m^2n^2 + 922mn + 2960 + \zeta(3)(480mn+2112) ] 
\frac{u^4}{432}  \right. \nonumber \\ 
&& \left. \,~~~~~-~ \left( 4[79mn + 1318 + 768\zeta(3)]u^3 \right.
\right. \nonumber \\ 
&& \left. \left. ~~~~~~~~~~~~-~ [555mn - 460(m+n) + 6836 + 4032\zeta(3)]u^2v 
\right. \right. \nonumber \\ 
&& \left. \left. ~~~~~~~~~~~~+~ 2[213mn - 358(m+n) + 1933 + 960\zeta(3)]uv^2 
\right. \right. \nonumber \\  
&& \left. \left. ~~~~~~~~~~~~-~ [121mn - 309(m+n) + 817 + 216\zeta(3)]v^3 
\right) \frac{(m-1)(n-1)v}{864} \right] \nonumber \\ 
&& +\, n^3 \! \left( [a_1 - a_2 - a_3 - a_4 - a_5 - a_6]u^5 + a_2u^4v 
+ a_3u^3v^2 + a_4u^2v^3 + a_5uv^4 + a_6v^5 \right) \nonumber \\
&& +~ O \left( u^6; \frac{1}{N^2} \right) 
\label{betu} 
\end{eqnarray} 
and 
\begin{eqnarray} 
\beta_v(u,v) &=& \frac{1}{2}(d-4)v ~+~ 2uv ~+~ \frac{1}{6}(m+n-8)v^2 ~-~ 
\frac{1}{18}(5mn+82)u^2v \nonumber \\ 
&& +~ \frac{1}{9}[5mn - 11(m+n) + 53]uv^2 ~-~ \frac{1}{36}[13mn - 35(m+n) + 99]
v^3 \nonumber \\ 
&& +~ \left[ \left( 52m^2n^2 - 57mn(m+n) - 2206mn - 111(m^2+n^2) + 4291(m+n) 
- 8084 \right. \right. \nonumber \\
&& \left. \left. ~~~~~~~-~ \zeta(3)( 1416mn - 3216(m+n) + 7392 ) \frac{}{}
\right) \frac{v^4}{864} \right. \nonumber \\ 
&& \left. ~~~~~~-~ \left( 39m^2n^2 - 35mn(m+n) - 1302mn - 36(m^2+n^2) 
+ 2401(m+n) \right. \right. \nonumber \\
&& \left. \left. ~~~~~~~~~~~~~-~ 5725 - \zeta(3)( 768mn - 1824(m+n) + 4896 ) 
\frac{}{} \right) \frac{v^3u}{216} \right. \nonumber \\ 
&& \left. ~~~~~~+~ \left( 78m^2n^2 - 35mn(m+n) - 2114mn + 3182(m+n) - 12520
\right. \right. \nonumber \\ 
&& \left. \left. ~~~~~~~~~~~~~-~ \zeta(3)( 1152mn - 2304(m+n) + 10368 ) 
\frac{}{} \right) \frac{u^2v^2}{432} \right. \nonumber \\   
&& \left. ~~~~~~-~ \left( 13m^2n^2 - 368mn - 3284 - \zeta(3)(192mn + 2688) 
\right) \frac{u^3v}{216} \right] \nonumber \\  
&& +~ N^3 \left[ (b_2 - b_3 - b_4 - b_5 - b_6)u^4v + b_3u^3v^2 + b_4u^2v^3 
+ b_5uv^4 + b_6v^5 \right] \nonumber \\
&& +~ O \left( u^6; \frac{1}{N^2} \right) 
\label{betv} 
\end{eqnarray} 
where we have rescaled the coupling constants by a numerical factor to ensure
the expressions are in the correct format for comparing with the Heisenberg
large $N$ value for $\omega$ and the ones we compute here. Also we have 
included the $d$-dependent terms as we are interested in the fixed point 
structure in $d$-dimensions. To assist with determining {\em new} information
on the four loop structure of both $\beta$-functions at $O(1/N)$ we have
introduced parameters, $\{a_i\}$ and $\{b_i\}$, for the coefficients of the
possible terms.  
\begin{figure}[ht] 
\epsfig{file=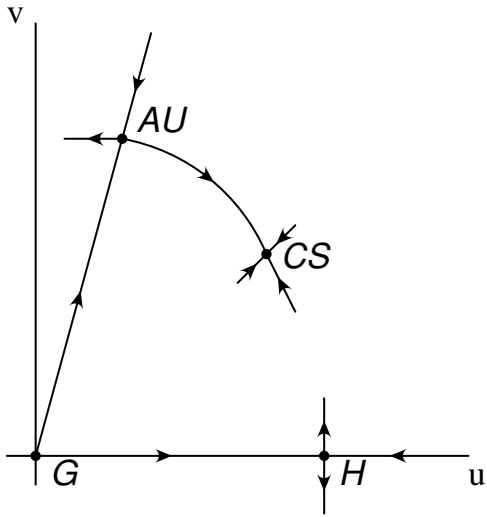,height=7cm} 
\vspace{0.5cm}
\caption{Renormalization group flow on the $(u,v)$ plane illustrating the
Gaussian ($G$), Heisenberg ($H$), stable chiral ($CS$) and unstable antichiral
($AU$) fixed points.} 
\end{figure}  

By examining the solutions to $\beta_u$ $=$ $0$ and $\beta_v$ $=$ $0$, 
\cite{8,11}, several fixed points emerge. First, there are the two obvious ones 
of the Gaussian fixed point, $u_c$ $=$ $0$, $v_c$ $=$ $0$, and the Heisenberg 
fixed point, $u_c$ $\neq$ $0$, $v_c$ $=$ $0$. For the latter point setting $v$ 
$=$ $0$ and $m$ $=$ $1$ in (\ref{betu}) one recovers the usual $O(N)$ symmetric
$\phi^4$ theory whose $\beta$-function is known at five loops in $\MSbar$ in 
four dimensions, \cite{2}. Indeed in our choice of parametrization of 
$\beta_v(u,v)$ at four loops we used this fact to restrict the function to be 
proportional to $v$. These two fixed points clearly lie on the axes of the 
$(u,v)$ coupling plane. However, for a range of values of $N$ and $m$ there are
two other fixed points which both have $u_c$ $\neq$ $0$ and $v_c$ $\neq$ $0$. 
One is known as CS and the other AU. The ultraviolet renormalization group flow
of the four fixed points in the $(u,v)$ plane is shown graphically in Figure 
$1$. The range of values for $N$ and $m$ for which such a renormalization group
flow is present has been detailed in \cite{8,11}, for example. However, for the 
purposes of the large $N$ calculation we will require the values of $u_c$ and 
$v_c$ to several order in $\epsilon$ and powers of $1/N$ where we take the
convention $d$ $=$ $4$ $-$ $2\epsilon$. In \cite{8,11} the 
explicit functions of $N$ and $m$ were presented at two loops. The full 
expressions at three loops can also be derived but are large, \cite{11}, and 
the {\em full} form is not necessary for our purposes. Indeed if we write 
\begin{eqnarray} 
u_c &=& \sum_{r=1}^\infty \left[ \frac{u_{r1}}{N} ~+~ \frac{u_{r2}}{N^2} 
\right] \epsilon^r ~+~ O \left( \frac{1}{N^3} \right) \nonumber \\ 
v_c &=& \sum_{r=1}^\infty \left[ \frac{v_{r1}}{N} ~+~ \frac{v_{r2}}{N^2} 
\right] \epsilon^r ~+~ \left( \frac{1}{N^3} \right) 
\end{eqnarray} 
for the $1/N$ expansion of the critical couplings in powers of $\epsilon$ the
four fixed points are determined to $O(\epsilon^4)$ and $O(1/N^2)$ as follows. 
For the Gaussian fixed point $u_{ri}$ $=$ $v_{ri}$ $=$ $0$. At the Heisenberg 
fixed point $u_{11}$~$=$~$6/m$, $u_{r1}$~$=$~$0$ for $r$~$\neq$~$1$, 
$u_{12}$~$=$~$-$~$48/m^2$, $u_{22}$~$=$~$108/m^2$, $u_{32}$~$=$~$-$~$99/m^2$,
$u_{42}$~$=$~$-$~$7776(a_1-a_2-a_3-a_4-a_5-a_6)/m^5$ and $v_{ri}$~$=$~$0$. At 
the stable fixed point, CS, 
\begin{eqnarray}  
u_{11} &=& 6 ~~,~~ u_{21} ~=~ 0 ~~,~~ u_{31} ~=~ 0 ~~,~~ u_{41} ~=~ 0 ~~,~~  
u_{12} ~=~ -~ 6m ~-~ 42 \nonumber \\
u_{22} &=& 18m ~+~ 90 ~~,~~ u_{32} ~=~ -~ \half ( 39m ~+~ 159 ) ~~,~~ 
u_{42} ~=~ -~ 7776 a_1 \nonumber \\  
v_{11} &=& 6  ~~,~~ v_{21} ~=~ 0  ~~,~~ v_{31} ~=~ 0  ~~,~~ v_{41} ~=~ 0  
v_{12} ~=~ -~ 6m ~-~ 24 ~~,~~ v_{22} ~=~ 18m ~+~ 54 \nonumber \\  
v_{32} &=& -~\half ( 39m ~+~ 99 ) ~~,~~ v_{42} ~=~ -~ 7776 b_2 ~. 
\end{eqnarray} 
Finally, at the unstable fixed point, AU,  
\begin{eqnarray} 
u_{11} &=& \frac{6(m-1)}{m} ~~,~~ u_{21} ~=~ 0 ~~,~~ u_{31} ~=~ 0 ~~,~~  
u_{12} ~=~ -~ 6m ~-~ 6 ~+~ \frac{108}{m} ~-~ \frac{96}{m^2} \nonumber \\ 
u_{22} &=& 18m ~+~ 12 ~-~ \frac{282}{m} ~+~ \frac{252}{m^2} ~~,~~
u_{32} ~=~ -~ \frac{39m}{2} ~-~ 66 ~+~ \frac{513}{2m} ~-~ \frac{171}{m^2} 
\nonumber \\ 
u_{42} &=& \frac{7776}{m^5} \left[ (m-1)^5 a_1 ~+~ (m-1)^4 a_2 ~+~ 
(2m-1)(m-1)^3 a_3 \right. \nonumber \\ 
&& \left. ~~~~~~~~+~ (3m^2-3m+1)(m-1)^2 a_4 ~+~ (2m^2-2m+1)(2m-1)(m-1) a_5 
\right. \nonumber \\
&& \left. ~~~~~~~~+~ (5m^4-10m^3+10m^2-5m+1) a_6 ~-~ 2(m-1)^5 b_2 ~-~ 
2(m-1)^4 b_3 \right. \nonumber \\
&& \left. ~~~~~~~~-~ 2(2m-1)(m-1)^3 b_4 ~-~ 2(3m^2-3m+1)(m-1)^2 b_5 \right.
\nonumber \\
&& \left. ~~~~~~~~-~ 2(2m^2-2m+1)(2m-1)(m-1) b_6 \right] \nonumber \\ 
v_{11} &=& 6 ~~,~~ v_{21} ~=~ 0 ~~,~~ v_{31} ~=~ 0 ~~,~~  
v_{12} ~=~ -~ 6m ~-~ 24 ~+~ \frac{72}{m} \nonumber \\  
v_{22} &=& 18m ~+~ 54 ~-~ \frac{204}{m} ~~,~~  
v_{32} ~=~ -~ \frac{39m}{2} ~-~ \frac{99}{2} ~+~ \frac{243}{m} \nonumber \\
v_{42} &=& -~ \frac{7776}{m^4} \left[ (m-1)^4 b_2 ~+~ (m-1)^3 b_3 ~+~ 
(2m-1)(m-1)^2 b_4 \right. \nonumber \\ 
&& \left. ~~~~~~~~~~~+~ (3m^2-3m+1)(m-1) b_5 ~+~ (2m^2-2m+1)(2m-1) b_6 
\right] ~. 
\end{eqnarray} 
These agree to two loops with the expressions given in \cite{8,11}. As we will 
be computing the critical exponents $\omega$ which relate to the critical slope
of the $\beta$-functions in large $N$ we can use these values to determine the 
$O(1/N)$ form of the critical exponents. As we are working with a two coupling 
model the stability exponents of each fixed point are related to the 
eigenvalues, $\lambda_I$, of the matrix of derivatives, $\Omega(u,v)$, 
evaluated at the appropriate fixed point where 
\begin{equation} 
\Omega(u,v) ~=~ \left( 
\begin{array}{cc} 
\frac{\partial \beta_u(u,v)}{\partial u} & 
\frac{\partial \beta_u(u,v)}{\partial v} \\ 
\frac{\partial \beta_v(u,v)}{\partial u} & 
\frac{\partial \beta_v(u,v)}{\partial v} \\ 
\end{array} 
\right) ~. 
\label{omeg} 
\end{equation} 
For the Gaussian case this is trivial and will not concern us here. For the
Heisenberg fixed point $\Omega(u,v)$ becomes triangular because $\beta_v(u,v)$
has no terms involving only $u$ at any order which implies
\begin{equation} 
\left. \frac{\partial \beta_v(u,v)}{\partial u} 
\right|^{\mbox{\footnotesize{Heis}}}_{u_c,v_c} ~=~ 0 ~. 
\end{equation}  
Thus, the critical point eigenvalues in the Heisenberg case 
are\footnote{It is worth stressing that our convention, $d$ $=$ $4$ $-$ 
$2\epsilon$, and the form we take for the $\beta$-functions, (\ref{betu}) and
(\ref{betv}), implies that for a stable direction the eigenexponent is 
$(4-d)/2$ $+$ $O(1/N)$. This differs from the standard result for the stability 
eigenexponent of $(4-d)$ $+$ $O(1/N)$ by a factor of $2$ which ought to be 
taken into account when comparing with other calculations. (See, for example, 
\cite{2}.)} 
\begin{eqnarray}
\lambda^{\mbox{\footnotesize{Heis}}}_+ &=& \epsilon ~-~ \frac{1}{mN} \left[ 
18 \epsilon^2 ~-~ 33 \epsilon^3 ~-~ \frac{3888}{m^3} [ a_1 - a_2 - a_3 - a_4 
- a_5 - a_6 ] \epsilon^4 ~+~ O(\epsilon^5) \right] \nonumber \\
&& +~ O \left( \frac{1}{N^2} \right) \nonumber \\ 
\lambda^{\mbox{\footnotesize{Heis}}}_- &=& -~ \epsilon ~+~ \frac{1}{mN} \left[ 
12 \epsilon - 10 \epsilon^2 ~-~ 13 \epsilon^3 ~+~ O(\epsilon^4) \right] ~+~ 
O \left( \frac{1}{N^2} \right) ~.  
\end{eqnarray} 
One of these corresponds to the stable direction in the renormalization group
flow, as indicated in Figure $1$, whilst the other corresponds to the unstable 
direction which will not concern us here. It is for the former for which the 
corrections at $O(1/N^2)$ have been computed in $d$-dimensions, \cite{26}, and 
we will record the $O(1/N)$ value later. For the remaining two fixed points the
critical matrix $\Omega(u_c,v_c)$ is not triangular. However, computing the 
eigenvalues at criticality of (\ref{omeg}) we find  
\begin{eqnarray} 
\lambda^{\mbox{\footnotesize{CS}}}_+ &=& \epsilon ~-~ \frac{1}{N} \left[ 
\frac{3(m^2+4m+7)\epsilon^2}{(m+1)} ~-~ \frac{(13m^4+72m^3+202m^2+216m+25)}
{2(m+1)^3} \epsilon^3 \right. \nonumber \\
&& \left. ~~~~~~~~~~~~-~ 4 \left[ 1944(m+1)^4 a_1 ~+~ 972(m-1)(m+1)^4 b_2 
\right. \right. \nonumber \\ 
&& \left. \left. ~~~~~~~~~~~~~~~~~~~-~ (2m^2+m-13)(m+2)(m-1) \right] 
\frac{\epsilon^4}{(m+1)^5} ~+~ O(\epsilon^5) \right] \, + \, O \left( 
\frac{1}{N^2} \right) \nonumber \\  
\lambda^{\mbox{\footnotesize{CS}}}_- &=& \epsilon ~-~ \frac{1}{N} \left[ 
6(m+1)\epsilon ~-~ \frac{(13m^2+26m+25)\epsilon^2}{(m+1)} \right. \nonumber \\
&& \left. ~~~~~~~~~~~~+~ \frac{(3m^4+12m^3+82m^2+116m-5)}{2(m+1)^3} 
\epsilon^3 \right. \nonumber \\ 
&& \left. ~~~~~~~~~~~~+~ 4 \left[ 324(2m-1)(m-1)(m+1)^4 a_1 + 324(m+1)^5 a_2
+ 648(m+1)^5 a_3 \right. \right. \nonumber \\ 
&& \left. \left. ~~~~~~~~~~~~~~~~~~~+~ 972(m+1)^5 a_4 + 1296(m+1)^5 a_5 
+ 1620(m+1)^5 a_6 \right. \right. \nonumber \\
&& \left. \left. ~~~~~~~~~~~~~~~~~~~-~ 648m(m-2)(m+1)^4 b_2 - 324(m+1)^5 b_3 
- 648(m+1)^5 b_4 \right. \right. \nonumber \\ 
&& \left. \left. ~~~~~~~~~~~~~~~~~~~-~ 972(m+1)^5 b_5 - 1296(m+1)^5 b_6 \right.
\right. \nonumber \\ 
&& \left. \left. ~~~~~~~~~~~~~~~~~~~-~ (2m^2+m-13)(m+2)(m-1) \right] 
\frac{\epsilon^4}{(m+1)^5} ~+~ O(\epsilon^5) \right] \, + \, O \left( 
\frac{1}{N^2} \right) \nonumber \\  
\lambda^{\mbox{\footnotesize{AU}}}_+ &=& \epsilon ~-~ \frac{1}{mN} \left[ 
(3m^2+9m-34)\epsilon^2 ~-~ \frac{1}{2}(13m^2+33m-162) \epsilon^3 \right. 
\nonumber \\
&& \left. ~~~~~~~~~~~~~~~-~ 3888 \left[ (m-1)^4 b_2 ~+~ (m-1)^3 b_3 ~+~ 
(2m-1)(m-1)^2 b_4 \right. \right. \nonumber \\ 
&& \left. \left. ~~~~~~~~~~~~~~~~~~~~+~ (3m^2-3m+1)(m-1) b_5 ~+~  
(2m-1)(2m^2-2m+1) b_6 \right] \frac{\epsilon^4}{m^3} \right. \nonumber \\
&& \left. ~~~~~~~~~~~~~~~+~ O(\epsilon^5) \right] ~+~ O \left( \frac{1}{N^2} 
\right) \nonumber \\  
\lambda^{\mbox{\footnotesize{AU}}}_- &=& -~ \epsilon ~+~ \frac{(m-1)(m+2)}{mN} 
\left[ 6\epsilon ~-~ 13\epsilon^2 ~+~ \frac{3}{2} \epsilon^3 ~+~ O(\epsilon^4) 
\right] ~+~ O \left( \frac{1}{N^2} \right) ~. 
\label{eigexpexp} 
\end{eqnarray} 
where the sign of the $O(\epsilon)$ terms relates to the stability property of
the fixed point when viewed from the ultraviolet renormalization group flow of 
Figure 1. For the exponents corresponding to the stable direction we have
included the $O(\epsilon^4)$ terms which depend on the unknown parameters of
the $O(1/N)$ four loop $\beta$-functions of (\ref{betu}) and (\ref{betv}) and
which will be constrained by our $O(1/N)$ critical exponents.  

\sect{Large $N$ formalism.} 
To compute the same critical eigenexponents in the large $N$ formalism in
$d$-dimensions one follows the programme of \cite{23,24} where the appropriate
Schwinger Dyson equations are analysed at the respective $d$-dimensional fixed
points of the theory. At this point the propagators scale asymptotically to
power law behaviour. In other words, in coordinate space, we have 
\begin{equation} 
\phi^{ai,bj}(x) ~=~ \delta^{ab} \delta^{ij}\phi(x) ~~~,~~~ \sigma(x) ~ \sim ~ 
\frac{B}{(x^2)^\beta} ~~~,~~~ \sigma^{ab,cd}_T(x) ~=~ X^{ab,cd} \sigma_T(x) 
\end{equation} 
where 
\begin{equation} 
\phi(x) ~ \sim ~ \frac{A}{(x^2)^\alpha} ~~~,~~~ \sigma_T(x) ~ \sim ~ 
\frac{C}{(x^2)^\gamma} 
\label{props} 
\end{equation} 
and 
\begin{equation} 
X^{ab,cd} ~=~ \frac{1}{2} \left( \delta^{ac} \delta^{bd} ~+~ \delta^{ad} 
\delta^{bc} ~-~ \frac{2}{m} \delta^{ab} \delta^{cd} \right) ~.  
\end{equation} 
This group theory factor satisfies the projector property 
\begin{equation}
X^{ab,pq} X^{pq,cd} ~=~ X^{ab,cd}
\end{equation} 
which implies that in labeling the internal indices on a diagram one only needs
to put indices on the $\phi$-field lines. The quantities $A$, $B$ and $C$ are 
the $x$-independent amplitudes of the fields and the exponents, $\alpha$, 
$\beta$ and $\gamma$ in our notation, are related to the wave function 
anomalous dimension $\eta$ by 
\begin{equation} 
\alpha ~=~ \mu ~-~ 1 ~+~ \half \eta ~~~,~~~ \beta ~=~ 2 ~-~ \eta ~-~ 
\chi ~~~,~~~ \gamma ~=~ 2 ~-~ \eta ~-~ \chi_T 
\end{equation} 
where $\chi$ and $\chi_T$ are the respective anomalous dimensions of the 
vertices involving $\sigma$ and $T^{ab}$ and $d$ $=$ $2\mu$. In \cite{11} 
$\eta$ was computed at $O(1/N^2)$ at both fixed points CS and AU. For the 
stable chiral one both the $\sigma$ and $T^{ab}$ fields propagate and couple. 
However, at the Heisenberg fixed point only the $\sigma$ field propagates since
clearly the $T^{ab}$ field is absent in the usual formulation of $\phi^4$ 
theory. Interestingly at the unstable antichiral fixed point the opposite 
situation emerges in that only the $T^{ab}$ field is present and the $\sigma$ 
field is omitted from the calculation of the large $N$ exponents, \cite{11}. 
Since the amplitudes appear in the calculations in the combinations $z$ $=$ 
$A^2B$ and $y$ $=$ $A^2C$ throughout and as they will be required for computing
our $\omega$ exponents we have determined their values at $O(1/N)$. For 
reference, at CS they are
\begin{equation} 
z_1 ~=~ -~ \frac{2\Gamma(2\mu-2)}{m\Gamma(2-\mu)\Gamma(\mu-2)} ~~~,~~~ 
y_1 ~=~ m z_1 
\end{equation} 
and at AU we have 
\begin{equation} 
z_1 ~=~ 0 ~~~,~~~ y_1 ~=~ -~ \frac{2\Gamma(2\mu-2)}{\Gamma(2-\mu) 
\Gamma(\mu-2)} ~. 
\end{equation} 
Moreover, for completeness the exponent $\eta$ at $O(1/N)$ is given by 
\begin{equation}
\eta ~=~ \sum_{i=1}^\infty \frac{\eta_i}{N^i} 
\end{equation} 
where 
\begin{eqnarray} 
\eta^{\mbox{\footnotesize{CS}}}_1 &=& -~ \frac{2(m+1)\Gamma(2\mu-2)} 
{\Gamma(\mu+1)\Gamma(\mu-1) \Gamma(\mu-2)\Gamma(2-\mu)} \nonumber \\ 
\eta^{\mbox{\footnotesize{AU}}}_1 &=& -~ \frac{2(m-1)(m+2)\Gamma(2\mu-2)} 
{m\Gamma(\mu+1)\Gamma(\mu-1) \Gamma(\mu-2)\Gamma(2-\mu)} ~.  
\end{eqnarray} 
Any subsequent exponent at either fixed point will be expressed in terms of 
their respective value for $\eta_1$. 

Since the exponents $\omega_I$ relate to corrections to scaling then to compute
them one considers corrections to the asymptotic scaling forms (\ref{props}), 
\cite{23,24,26}. In coordinate space we take  
\begin{eqnarray} 
\phi(x) & \sim & \frac{A}{(x^2)^\alpha} \left[ 1 ~+~ A^\prime (x^2)^\omega  
\right] ~~~,~~~ 
\sigma(x) ~ \sim ~ \frac{B}{(x^2)^\beta} \left[ 1 ~+~ B^\prime (x^2)^\omega 
\right] \nonumber \\ 
\sigma_T(x) & \sim & \frac{C}{(x^2)^\gamma} \left[ 1 ~+~ C^\prime (x^2)^\omega 
\right]
\label{propom} 
\end{eqnarray} 
where $A^\prime$, $B^\prime$ and $C^\prime$ are the $x$-independent correction
to scaling amplitudes whose values are not important here. In addition to 
(\ref{propom}) one requires the scaling form of the inverse propagators which
are determined by inverting the Fourier transform of (\ref{propom}) in
momentum space. Thus
\begin{eqnarray} 
\phi^{-1}(x) & \sim & \frac{p(\alpha)}{(x^2)^{2\mu-\alpha}A} \left[ 1 ~-~ 
q(\alpha,\omega) (x^2)^\omega A^\prime \right] \nonumber \\  
\sigma^{-1}(x) & \sim & \frac{p(\beta)}{(x^2)^{2\mu-\beta}B} \left[ 1 ~-~ 
q(\beta,\omega) (x^2)^\omega B^\prime \right] \nonumber \\  
\sigma^{-1}_T(x) & \sim & \frac{p(\gamma)}{(x^2)^{2\mu-\gamma}C} \left[ 1 ~-~ 
q(\gamma,\omega) (x^2)^\omega C^\prime \right] 
\end{eqnarray} 
where the functions $p(x)$ and $q(x,y)$ are defined by 
\begin{equation} 
p(x) ~=~ \frac{a(x-\mu)}{a(x)} ~~~,~~~ 
q(x,y) ~=~ \frac{a(x-y) a(x+y-\mu)}{a(x) a(x-\mu)} 
\end{equation}
with $a(x)$ $=$ $\Gamma(\mu-x)/\Gamma(x)$. Further, in our notation each 
exponent $\omega_I$ in the stable direction will have the $1/N$ expansion 
\begin{equation}
\omega ~=~ (\mu - 2) ~+~ \sum_{i=1}^\infty \frac{\omega_i}{N^i} 
\end{equation} 
and are therefore related to the eigenexponents of $\Omega(u_c,v_c)$ by 
$\omega_I$ $=$ $-$ $\lambda_I$ respectively in our conventions. Moreover, as 
noted before our choice of $\omega$ differs from the usual definition by a
factor of $(-1/2)$.  
\begin{figure}[ht] 
\epsfig{file=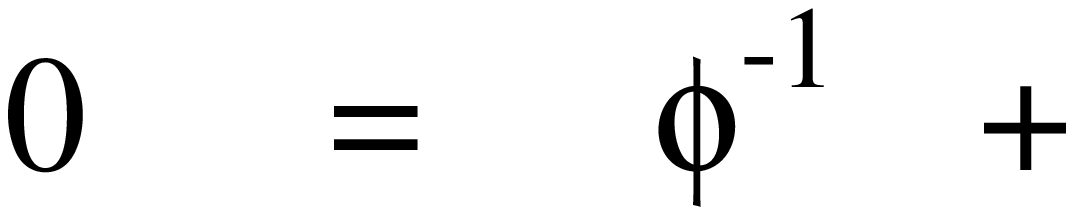,height=1.55cm} 
\vspace{0.5cm}
\caption{Schwinger Dyson equation for $\phi$ field at $O(1/N)$.} 
\end{figure}  

To solve for $\omega_1$ one substitutes the asymptotic scaling values into the
lines of the Schwinger Dyson equations of each of the fields. These are 
illustrated in Figures $2$ and $3$. In the latter we have only indicated the 
possible topologies that arise due to the nature of the expansion in powers
of $1/N$ and each wiggly line represents all possible combinations of $\sigma$ 
and $T^{ab}$ fields. If we ignore for the moment the two and three loop 
corrections in Figure $3$ we find the equations are represented by 
\begin{eqnarray}
0 &=& p(\alpha) \left[ 1 ~-~ q(\alpha,\omega) (x^2)^\omega A^\prime \right] ~+~
z \left[ 1 ~+~ \left( A^\prime + B^\prime \right) (x^2)^\omega \right] 
\nonumber \\
&& +~ \frac{(m-1)(m+2)y}{2m} \left[ 1 ~+~ \left( A^\prime + C^\prime \right) 
(x^2)^\omega \right] \nonumber \\ 
0 &=& p(\beta) \left[ 1 ~-~ q(\beta,\omega) (x^2)^\omega B^\prime \right] ~+~
\frac{1}{2} Nmz \left[ 1 ~+~ 2 (x^2)^\omega A^\prime \right] 
\nonumber \\
0 &=& p(\gamma) \left[ 1 ~-~ q(\gamma,\omega) (x^2)^\omega C^\prime \right] ~+~
\frac{1}{2} Ny \left[ 1 ~+~ 2 (x^2)^\omega A^\prime \right] 
\label{sdn} 
\end{eqnarray}
at the CS fixed point which we consider first for illustration. These equations
decouple on dimensional grounds into a set which determine $\eta$ at CS and a 
set involving $\omega$. The former lead to the values for $\eta_1$, $z_1$ and
$y_1$ quoted above. For the latter set of equations for consistency the
determinant of the $3$ $\times$ $3$ matrix defined with respect to the basis
vector $\{A^\prime,B^\prime,C^\prime\}$ has to vanish which naively leads to 
the equation 
\begin{equation}  
0 ~=~ \frac{(m-1)(m+2)}{m} y q(\beta,\omega) ~-~ q(\gamma,\omega) \left[
2z ~-~ (1 + q(\alpha,\omega)) q(\beta,\omega) \left[ z ~+~ 
\frac{(m-1)(m+2)}{2m} y \right] \right] \, .  
\label{sdnlo} 
\end{equation} 
This is similar to the consistency equation which determines the {\em exponent}
denoted by $\lambda$ in \cite{23,24} where the correction to scaling exponent 
has canonical dimension $\lambda_0$ $=$ $\mu$ $-$ $1$. However, for this value 
$q(\alpha,\lambda_0)$ with $\alpha$ $=$ $(\mu$ $-$ $1)$ or $2$ canonically is 
always an $O(1)$ quantity. By contrast, for the four dimensional model 
$q(\alpha,\omega_0)$ with $\omega_0$ $=$ $(\mu$ $-$ $2)$ and $\alpha$ $=$ 
$(\mu$ $-$ $1)$ or $2$ it has values which are $O(N)$ and $O(1/N)$ respectively
and therefore affects the consistency equation. For instance, whilst the 
product $q(\alpha,\omega) q(\beta,\omega) q(\gamma,\omega)$ in (\ref{sdnlo}) 
will be $O(1/N)$, this equation, (\ref{sdnlo}), would give an incorrect value 
for the exponents since the contributions in (\ref{sdnlo}) from $\sigma^{-1}$ 
and $\sigma_T^{-1}$ are of the same order in $1/N$ as those from the two and 
three loop correction graphs of Figure $3$ which we had naively omitted. Thus 
to determine $\omega_1$ correctly one must include the contributions from these
diagrams to the $B^\prime$ and $C^\prime$ parts of the Schwinger Dyson 
representation in the asymptotic approach to the fixed points. Hence, the 
$\sigma$ and $T^{ab}$ equations in (\ref{sdn}) are modified to 
\begin{eqnarray} 
0 &=& p(\beta) \left[ 1 ~-~ q(\beta,\omega) (x^2)^\omega B^\prime \right] ~+~
\frac{1}{2} Nmz \left[ 1 ~+~ 2 (x^2)^\omega A^\prime \right] \nonumber \\
&& +~ \left[ \frac{1}{2} Nm z^2 \Pi_1 B^\prime ~+~ \frac{1}{4} (m+2)(m-1) N yz 
\Pi_1 C^\prime \right. \nonumber \\ 
&& \left. ~~~~+~ N^2 m^2 z^3 \Pi_2 B^\prime ~+~ \frac{1}{2} 
(m+2)(m-1) N^2 y^2 z \Pi_2 C^\prime \right] (x^2)^\omega \nonumber \\ 
0 &=& p(\gamma) \left[ 1 ~-~ q(\gamma,\omega) (x^2)^\omega C^\prime \right] ~+~
\frac{1}{2} Ny \left[ 1 ~+~ 2 (x^2)^\omega A^\prime \right] \nonumber \\  
&& +~ \left[ \frac{1}{2} N yz \Pi_1 B^\prime ~+~ \frac{(m-2)}{4m} N y^2 
\Pi_1 C^\prime ~+~ N^2 y^2 z \Pi_2 B^\prime \right. \nonumber \\ 
&& \left. ~~~~+~ N^2 y^2 z \Pi_2 C^\prime ~+~ \frac{(m-2)(m+4)}{4m} N^2 y^3 
\Pi_2 C^\prime \right] (x^2)^\omega ~.  
\end{eqnarray}
In these equations to ensure the correct contribution to the $\omega$ Schwinger
Dyson equation after decoupling one correction term, $(x^2)^\omega$, is on one 
internal $\sigma$ or $T^{ab}$ line. There are corrections to the $\phi$ lines 
but these only contribute to $\omega_2$ after examining where they appear in 
the consistency determinant. This feature of having to include higher order 
graphs to obtain the correct $\omega_1$ is not peculiar to the CS fixed point 
as it already occurs at the Heisenberg point, \cite{26}. With these additional 
diagrams the correct consistency equation is given by setting the determinant 
of the matrix
\begin{equation}
\left[ 
\begin{array}{ccc}
(1+q(\alpha,\omega)) \left[ z + \frac{(m-1)(m+2)}{2m}y \right] & 
z &
\frac{(m-1)(m+2)}{2m}y \\ 
2 & 
q(\beta,\omega) + z\Pi_1 + 2mNz^2\Pi_2 &
\frac{(m-1)(m+2)}{2m}y ( \Pi_1 + 2Ny\Pi_2) \\
2 & 
z( \Pi_1 + 2Ny \Pi_2 ) &
\left[ q(\gamma,\omega) + \frac{(m-2)(m+4)}{2m}Ny^2 \Pi_2 \! \! \right. \\ 
&
&
\left. + \frac{(m-2)}{2m} y\Pi_1 + 2Nyz\Pi_2 \right] \\ 
\end{array}
\right] 
\label{matdetnt} 
\end{equation} 
to zero.  
\begin{figure}[ht] 
\epsfig{file=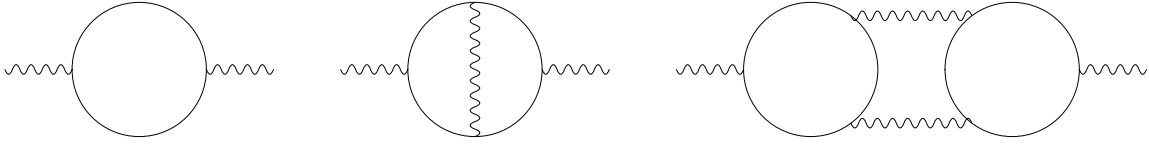,height=1.97cm} 
\vspace{0.5cm}
\caption{Basic topologies for the corrections to the $\sigma$ and $T^{ab}$ 
Schwinger Dyson equations to determine $\omega$ at $O(1/N)$.} 
\end{figure}  

Therefore, to solve for the respective $\omega$'s from (\ref{matdetnt}) the 
values of these additional diagrams must be computed. As the graphs are similar
to those used in the $O(N)$ $\phi^4$ calculation of $\omega$ if the group 
theory of each graph is suppressed we merely quote the $d$-dimensional {\em 
values} for the respective two and three loop graphs of Figure $3$. They are, 
ignoring symmetry factors, 
\begin{equation}
\Pi_1 ~=~ [\nu(2,\mu-1,\mu-1)]^2 ~~~,~~~ 
\Pi_2 ~=~ [\nu(2,\mu-1,\mu-1)]^2 \nu(1,2,2\mu-3) \nu(4-\mu,\mu-1,2\mu-3)  
\end{equation}
where $\nu(x,y,z)$ $=$ $a(x) a(y) a(z)$. These were calculated using the method
of uniqueness of \cite{28} which was extended from three dimensions to 
$d$-dimensions in \cite{23,24}. Futher, since the $O(1/N)$ values of 
$q(\beta,\omega)$ and $q(\gamma,\omega)$ now need to be included it transpires 
that the $O(1/N)$ expression for the vertex anomalous dimensions, $\chi$ and 
$\chi_T$, are required. This is due to 
\begin{equation} 
q(\beta,\omega) ~=~ \frac{a(4-\mu)\Gamma(\mu)}{a(2)a(2-\mu)} 
\frac{[\omega_1 - \eta_1 - \chi_1]}{N} ~+~ O \left( \frac{1}{N^2} \right) ~. 
\end{equation}  
We have computed each vertex anomalous dimension at each fixed point and find 
\begin{eqnarray} 
\chi^{\mbox{\footnotesize{CS}}}_1 &=& -~ \frac{\mu(4\mu-5) 
\eta^{\mbox{\footnotesize{CS}}}_1}{(\mu-2)} \nonumber \\ 
\chi^{\mbox{\footnotesize{CS}}}_{T,1} &=& -~ \frac{\mu[(2\mu-3)m + (4\mu-5)] 
\eta^{\mbox{\footnotesize{CS}}}_1}{(\mu-2)(m+1)} \nonumber \\ 
\chi^{\mbox{\footnotesize{AU}}}_{T,1} &=& -~ \frac{\mu(m-2)[(m+4)(2\mu-3) + 1] 
\eta^{\mbox{\footnotesize{AU}}}_1}{(m-1)(m+2)(\mu-2)} ~.  
\end{eqnarray} 
The expression for $\chi^{\mbox{\footnotesize{CS}}}_1$ is formally the same as
that for the $O(1/N)$ vertex dimension at the Heisenberg fixed point. For each  
exponent we have computed the value by applying the technique of \cite{29} of 
large $N$ critical point renormalization of $3$-point functions to each vertex
at the appropriate fixed point. For $\chi$ we have checked that the value 
agrees with that given by the scaling law which emerges from the theory which 
is in the same universality class as (\ref{lagcon}). This is believed to be the 
$O(N)$~$\times$~$O(m)$ two dimensional nonlinear sigma model where instead of
the term quadratic in $\sigma$ of (\ref{lagcon}) one has a term linear in
$\sigma$ where its coupling constant is related to the critical exponent $\nu$.
However, in such a model it is clear from group theory that there can be no 
linear term in $T^{ab}$ and therefore both expressions for $\chi_T$ cannot be 
derived from a scaling law but only direct calculation. 

With these values we can now determine the solution for the consistency 
equation for each fixed point. For CS since the matrix is $3$ $\times$ $3$ two 
values for $\omega$ emerge. These are 
\begin{eqnarray}
\omega^{\mbox{\footnotesize{CS}}}_{\pm} &=& (\mu-2) ~+~ 
\frac{(2\mu-1)\eta^{\mbox{\footnotesize{CS}}}_1}{2(m+1)(\mu-2)N} \left[ 
m(\mu-1)(\mu-4) + (2\mu^2 - 7\mu + 4) \right. \nonumber \\ 
&& \left. ~~~~~~ \pm~ \mu \left[ (m^2-1)(\mu-1)^2 + 2(m-1)(2\mu-3)(\mu-1) 
+ (5\mu-8)^2 \right]^{\half} \right] ~+~ O \left( \frac{1}{N^2} \right) 
\nonumber \\ 
\label{omlncs} 
\end{eqnarray}  
where the $\pm$ subscript refers to the sign in front of the discriminant. Both
are perfectly acceptable solutions since one corresponds to one stable
direction at CS and the other to the eigenexponent from the second direction.

For AU the derivation of the consistency equation follows the same pattern as
that for CS in that two and three loop graphs have also to be included due to
the large $N$ counting. The consistency equation itself can be deduced from 
(\ref{matdetnt}) by deleting the second row and column from the matrix and 
setting $z$ $=$ $0$ in the remaining entries. Since only the $T^{ab}$ field is 
present only one value for $\omega$ emerges which is 
\begin{equation}
\omega^{\mbox{\footnotesize{AU}}}_+ ~=~ (\mu-2) ~-~ \left[ 2\mu^2 - 3\mu 
- 1 ~+~ \frac{\mu (m-2) [ 2\mu-5 - 2(m+4)(2\mu-3) ]}{(m-1)(m+2)} \right] 
\frac{\eta^{\mbox{\footnotesize{AU}}}_1}{N} ~+~ O \left( \frac{1}{N^2} \right)  
\label{omlnau} 
\end{equation}  
which corresponds to the eigenexponent of the stable direction at AU. For 
completeness we note that at the Heisenberg fixed point for the stable 
direction we have, \cite{27,26},   
\begin{equation}  
\omega^{\mbox{\footnotesize{Heis}}}_+ ~=~ (\mu-2) ~-~ \frac{4(2\mu-1)^2 
\Gamma(2\mu-2)}{\Gamma(2-\mu)\Gamma(\mu-1)\Gamma(\mu-2) \Gamma(\mu+1)mN} ~+~ 
O \left( \frac{1}{N^2} \right) 
\label{omlnheis} 
\end{equation}  
which is deduced by deleting the third row and column of (\ref{matdetnt}) and
setting $y$ $=$ $0$.  

In order to check the correctness of each result one can set $d$ $=$ $4$ $-$
$2\epsilon$, or $\mu$ $=$ $2$ $-$ $\epsilon$, and expand each exponent to 
$O(\epsilon^3)$ and compare with the values of explicit perturbation theory, 
(\ref{eigexpexp}). We find total agreement for all three exponents and 
therefore regard (\ref{omlncs}) and (\ref{omlnau}) as correct. Indeed equipped 
with (\ref{omlncs}) and (\ref{omlnau}) we can now reverse the check argument 
and use the information contained in the $d$-dimensional exponents to determine
constraints on the structure of the {\em four} loop $\MSbar$ corrections to 
(\ref{betu}) and (\ref{betv}). First, we record the $\epsilon$ expansion of 
each of the exponents in the stable directions which are consistent with 
(\ref{eigexpexp}).  From (\ref{omlncs}), (\ref{omlnau}) and (\ref{omlnheis}) we
have  
\begin{eqnarray} 
\omega^{\mbox{\footnotesize{CS}}}_{+,1} &=& \left[ \frac{3(m^2+4m+7)}{(m+1)} 
\epsilon^2 ~-~ \frac{(13m^4+72m^3+202m^2+216m+25)}{2(m+1)^3} \epsilon^3 \right.
\nonumber \\ 
&& \left. ~+~ \frac{(3m^6+10m^5+21m^4-36m^3-387m^2-326m+395)}{4(m+1)^5} 
\epsilon^4 ~+~ O(\epsilon^5) \right] \nonumber \\
\omega^{\mbox{\footnotesize{CS}}}_{-,1} &=& \left[ 6(m+1)\epsilon ~-~ 
\frac{(13m^2+26m+25)}{(m+1)} \epsilon^2 ~+~ 
\frac{(3m^4+12m^3+82m^2+116m-5)}{2(m+1)^3} \epsilon^3 \right. \nonumber \\
&& \left. ~+~ \frac{[ 48(m+1)^6 \zeta(3) + 3m^6 + 18m^5 + 29m^4 + 76m^3 
+ 397m^2 + 322m - 397]}{4(m+1)^5} \epsilon^4 \right. \nonumber \\ 
&& \left. ~+~ O(\epsilon^5) \right] \nonumber \\ 
\omega^{\mbox{\footnotesize{AU}}}_{+,1} &=& \frac{1}{m} \left[ (3m^2+9m-34) 
\epsilon^2 ~-~ \frac{1}{2} (13m^2+33m-162) \epsilon^3 ~+~ \frac{1}{4} 
(3m^2-5m-54) \epsilon^4 \right. \nonumber \\
&& \left. ~~~~+~ O(\epsilon^5) \right] \nonumber \\  
\omega^{\mbox{\footnotesize{Heis}}}_{+,1} &=& \frac{1}{m} \left[ 
18 \epsilon^2 ~-~ 33 \epsilon^3 ~-~ \frac{5}{2} \epsilon^4 ~+~ O(\epsilon^5) 
\right] ~.  
\end{eqnarray} 
Next we compare the $O(\epsilon^4)$ coefficients with those of the explicit
perturbative expansion which have been parametrized by $\{a_i\}$ and $\{b_i\}$.
From the Heisenberg exponent we have 
\begin{equation} 
a_1 ~-~ a_2 ~-~ a_3 ~-~ a_4 ~-~ a_5 ~-~ a_6 ~=~ \frac{5m^2}{7776} 
\end{equation}
which agrees with the explicit four loop $\phi^4$ computation when one sets
$m$ $=$ $1$ and determines the $O(1/N)$ coefficient of the $u^5$ term of
$\beta_u(u,v)$. From the CS fixed point we have 
\begin{equation} 
2 a_1 ~+~ (m-1) b_2 ~=~ -~ \frac{(3m+7)(m-3)}{15552} 
\end{equation} 
and 
\begin{eqnarray} 
&& (m+1) [ a_2 ~+~ 2 a_3 ~+~ 3 a_4 ~+~ 4 a_5 ~+~ 5 a_6 ~-~ b_3 ~-~ 2 b_4 ~-~ 
3 b_5 ~-~ 4 b_6 ] \nonumber \\  
&& +~ (2m-1)(m-1) a_1 ~-~ 2m(m-2) b_2 \nonumber \\
&& =~ \frac{\zeta(3)}{108(m+1)^3} ~+~ \frac{(3m^2+6m+19)}{5184(m+1)^5} ~.  
\end{eqnarray} 
The final constraint arises from the stable direction at the AU fixed point
which gives 
\begin{eqnarray}
&& (m-1)^4 b_2 ~+~ (m-1)^3 b_3 ~+~ (2m-1)(m-1)^2 b_4 ~+~ (3m^2-3m+1)(m-1) b_5 
\nonumber \\ 
&& +~ (2m-1)(2m^2-2m+1) b_6 \nonumber \\
&& =~ -~ \frac{[3m^2-5m-54]}{15552} ~.  
\end{eqnarray} 

One can now deduce the values for the exponents in the four stable directions 
in three dimensions. These are, with our conventions,  
\begin{eqnarray} 
\omega^{\mbox{\footnotesize{CS}}}_+ &=& -~ \frac{1}{2} ~+~ 
\frac{4(m+4)}{3\pi^2 N} ~+~ O \left( \frac{1}{N^2} \right) \nonumber \\  
\omega^{\mbox{\footnotesize{CS}}}_- &=& -~ \frac{1}{2} ~+~ 
\frac{16(m+1)}{3\pi^2 N} ~+~ O \left( \frac{1}{N^2} \right) \nonumber \\
\omega^{\mbox{\footnotesize{AU}}}_+ &=& -~ \frac{1}{2} ~+~ 
\frac{4[m^2+4m-8]}{3\pi^2 mN} ~+~ O \left( \frac{1}{N^2} \right) \nonumber \\ 
\omega^{\mbox{\footnotesize{Heis}}}_+ &=& -~ \frac{1}{2} ~+~ 
\frac{32}{3\pi^2 mN} ~+~ O \left( \frac{1}{N^2} \right) ~.  
\label{om3exp} 
\end{eqnarray} 
For CS the corrections both have the same sign and interestingly neither
involves a square root which appears in the $d$-dimensional expression. For 
the specific case of $m$ $=$ $2$ we have from (\ref{om3exp})  
\begin{eqnarray} 
\left. \omega^{\mbox{\footnotesize{CS}}}_+ \right|_{m=2} &=& -~ \frac{1}{2} ~+~ 
\frac{8}{\pi^2 N} ~+~ O \left( \frac{1}{N^2} \right) \nonumber \\  
\left. \omega^{\mbox{\footnotesize{CS}}}_- \right|_{m=2} &=& -~ \frac{1}{2} ~+~ 
\frac{16}{\pi^2 N} ~+~ O \left( \frac{1}{N^2} \right) \nonumber \\
\left. \omega^{\mbox{\footnotesize{AU}}}_+ \right|_{m=2} &=& -~ \frac{1}{2} ~+~ 
\frac{8}{3\pi^2 N} ~+~ O \left( \frac{1}{N^2} \right) \nonumber \\ 
\left. \omega^{\mbox{\footnotesize{Heis}}}_+ \right|_{m=2} &=& -~ 
\frac{1}{2} ~+~ \frac{16}{3\pi^2 N} ~+~ O \left( \frac{1}{N^2} \right) ~.  
\label{omm_2} 
\end{eqnarray} 
With these values we can comment on the possible breakdown of stabilty in this
model. In our computations so far have relied on the fact that the stability 
picture for the fixed points represented in Figure 1 is valid for the full
range of $N$ in the large $N$ expansion. However, it has been suggested that 
for certain values of $N$ this scenario may be different and that the 
underlying assumption of the existence of a second order phase transition, 
around which the large $N$ critical point formalism is built, could break down.
Indeed there is some controversy in this model in three dimensions about the
existence of CS and whether the phase transition is first or second order for 
relatively low values of $N$. Moreover, the precise value of $N$ where the 
order changes has not been determined {\em consistently} from different 
methods. (A recent review is given in \cite{17}.) For instance, a value has 
been found for this critical value of $N$ by using $d$~$=$~$4$~$-$~$2\epsilon$ 
dimensional perturbation theory and it was extended to three dimensions using 
standard resummation techniques, \cite{30}, giving $N_c$ $=$ $3.39$. By 
contrast, Monte Carlo methods and another $\epsilon$-expansion extraction have 
suggested $N_c$ $<$ $2$, \cite{6,7,8}. With the large $N$ corrections 
(\ref{omm_2}) we can naively examine the range of values of $N$ for which 
either CS stability exponent changes sign. For 
$\left. \omega^{\mbox{\footnotesize{CS}}}_- \right|_{m=2}$ this will occur
when $N_c$ $=$ $3.24$ whilst  
$\left. \omega^{\mbox{\footnotesize{CS}}}_+ \right|_{m=2}$ changes sign when
$N_c$~$=$~$1.62$. The former would suggest a critical $N$ in a similar range
to that of \cite{30}. However, these remarks ought to be qualified with various
observations. First, we have only computed the $O(1/N)$ correction to stability
where $N$ is assumed to be large. Therefore, one has to ask whether the 
approximation will still be valid for such a low value of $N$. Second, the
nature of the large $N$ expansion is a reordering of perturbation theory such
that a certain class of diagrams are summed first. Therefore, if one could
compute to all orders one would reproduce ordinary perturbation theory and so
obtaining a value for $N_c$ in three dimensions which is not inconsistent with
the resummed value determined from several loop orders in ordinary perturbation
theory would seem only to reinforce that particular value. In other words if 
non-perturbative effects become significant at CS for low values of $N$ to
affect the precise location of $N_c$ these will have been omitted in 
perturbation theory. Third, for our value of $N_c$ we have naively assumed the 
large $N$ series is convergent and therefore that our simple assumption that
when the $O(1/N)$ correction exceeds $1/2$ the character of the fixed point 
changes is valid. Only a higher order calculation would resolve this.  

\sect{Discussion.} 
We have computed the correction to scaling exponents $\omega$ in all the stable
directions of the Landau-Ginzburg-Wilson model with $O(N)$ $\times$ $O(m)$ 
symmetry at $O(1/N)$ in $d$-dimensions. This allows one to extract information 
in {\em all} the available large $N$ exponents of \cite{11} at $O(1/N)$ in 
relation to the four dimensional theory in the same way that the two 
dimensional critical slope information contained in the exponent $\nu$ does for
the underlying two dimensional theory. It is also worth stressing again, 
\cite{11}, that from the point of view of the large $N$ formalism a consistent 
picture emerges in terms of the active fields of the theory formulated in terms
of the auxiliary fields $\sigma$ and $T^{ab}$ of (\ref{lagcon}). At the 
Heisenberg and AU fixed points only $\sigma$ and $T^{ab}$ respectively 
propagate which corresponds in the large $N$ formalism developed here to one 
stable direction and hence only one eigenexponent emerged. However, at the only
fully stable fixed point both fields are relevant leading to two independent 
stability exponents. This is a natural way to picture this particular model 
which we assume persists to higher orders in large $N$. Moreover, our 
consistency with three loop $\MSbar$ perturbation theory provides an important
internal cross check on the values of quantities, such as the vertex anomalous 
dimensions, which we had to compute en route to our expressions for $\omega_i$.
Further, the information contained within the expressions (\ref{omlncs}), 
(\ref{omlnau}) and (\ref{omlnheis}) will provide important checks on any future
explicit four loop $\MSbar$ perturbative calculations which would improve the 
accuracy of the numerical estimates deduced from (\ref{betu}), (\ref{betv}) and
other renormalization group functions. Such four loop calculations are 
certainly viable since five loop results are available in $\MSbar$ in ordinary 
$O(N)$ $\phi^4$ theory. For example, the integration routines for the four loop
Feynman diagrams have already been constructed. However, one can also attack 
this problem from the large $N$ point of view. For instance, we have 
demonstrated the elegance of the formalism to produce the critical 
eigenexponents at $O(1/N)$. However, this machinery has already been extended 
in \cite{26} to compute $\omega_2$ in the Heisenberg model. Therefore, we would
expect there to be no serious obstacles to extending the present computation to 
find $\omega^{\mbox{\footnotesize{CS}}}_2$ and 
$\omega^{\mbox{\footnotesize{AU}}}_2$. For example, the values of the
underlying three, four, five and six loop Feynman diagrams which are analogous
to the corrections of Figure $3$ for the $O(1/N^2)$ calculation have been
determined. We hope to return to this in a future article.

\end{document}